\documentclass[prd,aps,10pt,showkeys,nofootinbib,showpacs,twocolumn]{revtex4-1}

\usepackage{amsmath,amssymb,amsfonts,amsthm}
\usepackage{amsbsy} 
\usepackage{epsfig}
\usepackage{latexsym}
\usepackage{accents}
\input amssym.def
\input amssym.tex

\usepackage{hyperref}

\newcommand{\oarX}[1]{\href{http://arxiv.org/abs/#1}{{\ttfamily #1}}}
\newcommand{\arX}[1]{\href{http://arxiv.org/abs/#1}{{\ttfamily arXiv:#1}}}
\newcommand{\doin}[2]{\href{http://dx.doi.org/#1}{#2}}

\newcommand{\utilde}[1]{ \underaccent{\tilde}{#1} }

\def\barr{\begin{array}}
\def\earr{\end{array}}
\def\half{\frac{1}{2}}
\def\ben{\begin{equation}}
\def\een{\end{equation}}
\def\bs{\begin{subequations}}
\def\es{\end{subequations}}
\def\bena{\begin{eqnarray}}
\def\eena{\end{eqnarray}}

\begin{document}

\title{Canonical analysis of unimodular Pleba\'nski gravity}

\author{Steffen Gielen}
\affiliation{School of Mathematical and Physical Sciences, University of Sheffield, Hicks Building, Hounsfield Road, Sheffield S3 7RH, United Kingdom}
\email{s.c.gielen@sheffield.ac.uk}
\email{enash4@sheffield.ac.uk}
\author{Elliot Nash}
\affiliation{School of Mathematical and Physical Sciences, University of Sheffield, Hicks Building, Hounsfield Road, Sheffield S3 7RH, United Kingdom}
\date{\today}


\begin{abstract}

We present the canonical analysis of different versions of unimodular gravity defined in the Pleba\'nski formalism, based on a (generally complex) $SO(3)$ spin connection and set of (self-dual) two-forms. As in the metric formulation of unimodular gravity, one can study either a theory with fixed volume form or work in a parametrised formalism in which the cosmological constant  becomes a dynamical field, constrained to be constant by the field equations. In the first case, the Hamiltonian density contains a part which is not constrained to vanish, but rather constrained to be constant, again as in the metric formulation. We also discuss reality conditions and challenges in extracting Lorentzian solutions. 
\end{abstract}


\maketitle

\section{Introduction}

General relativity can be defined in terms of a number of different actions and using different dynamical variables. One can work with a spacetime metric as the field encoding spacetime curvature, or use an independent connection; one can introduce an additional local gauge symmetry and work with a tetrad and spin connection; or one can express the dynamics of gravity in terms of torsion or non-metricity. Many of these formulations are reviewed in Ref.~\cite{Krasnovbook}. The (chiral) Pleba\'nski formulation takes a slightly different viewpoint from other connection-based approaches by encoding the metric information not in a tetrad but in a set of complex 2-forms \cite{Plebanski}. The formulation is fundamentally complex since both the 2-forms and the $SO(3,\mathbb{C})$ spin connection live in the self-dual part (under Hodge duality) in the complexified algebra $\mathfrak{so}(3,1)_{\mathbb{C}}$; since the Hodge dual squares to $-1$ in Lorentzian signature a complexification is required.  The identification of solutions that correspond to Lorentzian signature solutions to general relativity then requires adding reality conditions to the theory. In Euclidean signature, one can work with self-dual real $SO(3)$ variables instead. 

The Hamiltonian formulation of the chiral Pleba\'nski theory is equivalent to the complex Ashtekar formulation of Hamiltonian general relativity \cite{selfdual2forms}, itself the basis for canonical quantisation in loop quantum gravity \cite{Ashtekar}. The result is a simple and elegant version of Hamiltonian general relativity with polynomial constraints and structural similarities to Yang--Mills gauge theory. Again, one needs to add reality conditions which has proven to be very difficult in the quantum theory, so that alternative (non-chiral) real-valued formulations of Pleba\'nski gravity were introduced. The canonical analysis of these formulations is also known, and more involved \cite{RealPlebanski}. In this article we focus on chiral complex formulations, and the Hamiltonian analysis will be more straightforward; we will discuss the issue of reality conditions.

Our focus in this article is a canonical analysis of unimodular gravity in the Pleba\'nski formulation. {\em Unimodular gravity} is a term that can apply to a number of different versions of general relativity based on different actions and symmetry principles (see, e.g., Refs.~\cite{UnimRefs,Moreorless}), which are locally equivalent to general relativity but promote the cosmological constant from a constant of Nature to a global integration constant. Hence, these theories have one global degree of freedom in addition to those of standard general relativity. This can be achieved by restricting the diffeomorphism symmetry of general relativity by fixing a volume form \cite{RestUG}, or by introducing additional fields representing a dynamical cosmological constant as done by Henneaux and Teitelboim in Ref.~\cite{HenneauxTeitelboim}. Both approaches can be implemented in the Pleba\'nski formalism as shown in Ref.~\cite{UnimPleb}; a version based on the Henneaux--Teitelboim approach was already discussed in Ref.~\cite{Smolin}. We will present the canonical analysis of both approaches, including some results that can already be found in Ref.~\cite{Smolin}, while also discussing additional issues of imposing reality conditions and obtaining a metric of different types of Lorentzian signature. Most expressions will apply to both Lorentzian and Euclidean cases.

We start in Sec.~\ref{unimreview} by presenting the unimodular formulations of Pleba\'nski gravity proposed in Ref.~\cite{UnimPleb}. In Sec.~\ref{canonsec1} we will first give an extended summary of the canonical analysis of standard Pleba\'nski gravity, which will then facilitate the extension to various unimodular versions of theory given in Sec.~\ref{canonsec2}. We conclude in Sec.~\ref{conclus}. In Appendix~\ref{appendix} we derive the transformation behaviour of Lagrange multipliers under gauge transformations, based on invariance of the action.

\section{Unimodular Pleba\'nski gravity}
\label{unimreview}

The starting point for our discussions is the chiral Pleba\'nski formulation of general relativity in four spacetime dimensions. The action can be written as
\begin{align}
S_{{\rm Pl}}[A,\Sigma,M,\omega]=&\,\frac{1}{8\sqrt{\sigma}\,\pi G}\int \bigl[\Sigma_i\wedge F^i - \frac{1}{2}M^{ij}\Sigma_i\wedge \Sigma_j \nonumber
\\&+ \frac{1}{2}\omega\left({\rm tr}\, M-\Lambda\right)\bigr]\,,
\label{Plebanskiaction}
\end{align}
where $A$ is an $SO(3)$ connection, $F$ is the curvature 2-form of $A$, $\Sigma^i$ are 2-forms, $M^{ij}$ is a symmetric matrix field, and $\omega$ is a 4-form. At this point $\Lambda$ is seen as a fixed parameter corresponding to the cosmological constant. In the Euclidean case, all variables are real-valued and $\sigma=1$; in the Lorentzian case all variables are complex valued and $\sigma=-1$ so that $\sqrt{\sigma}={\rm i}$, and reality conditions are needed to identify Lorentzian solutions. 

The field equations
\begin{align}
\Sigma^i\wedge\Sigma^j = &\, \delta^{ij}\omega\,,\qquad D_A\Sigma^i=0\,,\nonumber
\\ F^i = & \,M^{ij}\Sigma_j\,,\qquad {\rm tr}\,M=\Lambda
\label{fieldeq}
\end{align}
encode the dynamical content of general relativity: the condition $\Sigma^i\wedge\Sigma^j = \delta^{ij}\omega$ means that $\Sigma^i$ can be written in terms of a tetrad $e^I$, with the general solution in the complex case given by \cite{Krasnovbook,selfdual2forms,UnimPleb}
\ben
\Sigma^i = {\rm i}\,e^0\wedge e^i - \half {\epsilon^i}_{jk} e^j \wedge e^k\,.
\label{tetracond}
\een
Then the second equation means that the connection is torsion-free. The equations in the second line of Eq.~(\ref{fieldeq}) encode the Einstein condition $R_{\mu\nu}\propto g_{\mu\nu}$ and the trace part of the Einstein equations, $R=4\Lambda$. More details on all these identifications can be found in Ref.~\cite{Krasnovbook}. Again, we stress that in the complex case Lorentzian solutions need to be identified by imposing separate reality conditions. (Internal $SO(3)$ indices $i,j$ can be raised and lowered with the Kronecker delta $\delta_{ij}$ so their positioning is not important.)

An important insight in the metric formulation of general relativity is that the trace equation $R=4\Lambda$ is almost redundant: from the trace-free part of the Einstein equations and the Bianchi identities it follows that $R$ must be a spacetime constant, which one may then identify with what is usually thought of as $\Lambda$ \cite{Tracefree}. The same goes through in the Pleba\'nski formulation; assuming the other three equations, the trace of $M$ must be an arbitrary constant. Hence, one can seek modified action principles in which the trace equation is not separately imposed. One finds a preferred-volume version
\begin{align}
S_{{\rm PV}}[A,\Sigma,M;\omega_0]=&\,\frac{1}{8\sqrt{\sigma}\,\pi G}\int \bigl[\Sigma_i\wedge F^i - \frac{1}{2}M^{ij}\Sigma_i\wedge \Sigma_j \nonumber
\\&+ \frac{1}{2}\omega_0\,{\rm tr}\, M\bigr]\,,
\label{preferredvolume}
\end{align}
where $\omega_0$ is a background volume form and not dynamical, and a parametrised Henneaux--Teitelboim version
\begin{align}
S_{{\rm HT}}[A,\Sigma,M,T]=&\,\frac{1}{8\sqrt{\sigma}\,\pi G}\int \bigl[\Sigma_i\wedge F^i - \frac{1}{2}M^{ij}\Sigma_i\wedge \Sigma_j \nonumber
\\&+ {\rm d}T\,{\rm tr}\, M\bigr]\,,
\label{plebanskiht}
\end{align}
which now depends on a 3-form field $T$. Notice that neither of these actions contains a parameter $\Lambda$, as expected in unimodular formulations.

In the preferred-volume theory we now no longer impose the trace equation, but show that it emerges as a consequence of the other three; in the parametrised version we have a new equation ${\rm d}\,{\rm tr}\,M=0$ setting ${\rm tr}\,M$ to an arbitrary constant, and the volume form is now proportional to ${\rm d}T$ for some $T$. This last property allows constructing a global ``volume time'' from $T$ as we will detail below. 

As in the conventional Pleba\'nski approach, one can obtain alternative formulations by ``integrating out'' fields, starting by replacing $\Sigma^i=(M^{-1})_{ij}F^j$ and then also eliminating the matrix field $M$ to obtain ``pure connection'' formulations of these theories, in analogy with what is done for general relativity \cite{PureConnection}. These constructions are detailed in Ref.~\cite{UnimPleb} and we will comment on them below.

\section{Canonical Pleba\'nski gravity}
\label{canonsec1}

We start by presenting a canonical analysis of chiral Pleba\'nski gravity defined by the action (\ref{Plebanskiaction}). While this has been discussed in the literature previously, the purpose of our discussion here is to introduce our conventions which will then also be used when moving to unimodular formulations. We also aim to be particularly explicit in interpreting the gauge transformations generated by constraints, and in discussing reality conditions and the role of a spacetime metric.

As usual, we assume that spacetime has topology $\mathcal{M}=\mathbb{R}\times\mathcal{S}$ where $\mathcal{S}$ is either compact without boundary or suitable fall-off conditions are imposed on the fields. We introduce coordinates $(t,x^a)$ and decompose the connection and 2-forms as
\begin{align}
A^i = &\, A^i_0\,{\rm d}t + A^i_a\,{\rm d}x^a\,,\nonumber
\\\Sigma^i = &\, \Sigma^i_{0a}\,{\rm d}t\wedge{\rm d}x^a + \half \Sigma^i_{ab}\,{\rm d}x^a\wedge{\rm d}x^b\,.
\label{sigmadecompose}
\end{align}
In addition, the 4-form $\omega$ can be written as $\omega=-2\tilde{N}\,{\rm d}^4 x$ for some scalar density $\tilde{N}$.

The spatial components $\Sigma^i_{ab}$ define a {\em densitised triad}
\ben
\tilde{E}^{a}_i = \half \tilde\epsilon^{abc}\Sigma_{ibc}
\label{denstriad}
\een
where $\tilde\epsilon^{abc}$ is the usual totally antisymmetric tensor density of weight $+1$ with $\tilde\epsilon^{123}=1$. Here and in the following we use tildes to indicate density weights. $\tilde{E}^{a}_i$ is the densitised triad appearing in the Ashtekar formulation of general relativity \cite{Ashtekar}, as already identified in Ref.~\cite{selfdual2forms}. Assuming that the matrix $\tilde{E}^a_i$ is invertible, it has an inverse
\ben
\utilde{E}^i_a=\frac{\utilde{\epsilon}_{abc}\epsilon^{ijk}\tilde{E}_j^b \tilde{E}^c_k}{2\det\tilde{E}}
\label{invtriad}
\een
where $\det\tilde{E}$ is the determinant of the matrix with entries $\tilde{E}^a_i$. $\utilde{E}^i_a$ defines a densitised co-triad, a Lie algebra valued covector density of weight $-1$ (as indicated by the tilde underneath). 

As the last step in introducing canonical variables, it is convenient to parametrise the components $\Sigma^i_{0a}$ as
\ben
\Sigma^i_{0a} = -\utilde{\epsilon}_{abc} V^b \tilde{E}^{ic} - \tilde\varphi^{ij}\utilde{E}_{ja}\,,
\een
where $\tilde\varphi^{ij}$ is a symmetric matrix field of density weight $+1$. This parametrisation is non-degenerate, i.e., $\Sigma^i_{0a}=0$ only when $V^b=\tilde\varphi^{ij}=0$. We also write the field $A^i_0$, which appears without time derivatives in the action, as $A^i_0=\alpha^i+V^a A^i_a$. With all these definitions, one finds
\begin{align}
&\Sigma_i\wedge\left(F^i - \frac{1}{2}M^{ij} \Sigma_j\right) \nonumber
\\ = &\,{\rm d}^4 x\,\bigl[\dot{A}^i_a\tilde{E}_i^a + \alpha^i\,D_a\tilde{E}^a_i-V^a\left(F^i_{ab}\tilde{E}^b_i-A^i_a D_b\tilde{E}^b_i\right)\nonumber
\\ & -\tilde\varphi_{ij}\left(\frac{F^i_{ab}\epsilon^{jkl}\tilde{E}^a_k\tilde{E}^b_l}{2\det\tilde{E}}-M^{ij}\right)-\partial_a\left(A_0^i\tilde{E}^a_i\right)\bigr]\,,
\label{decomposition}
\end{align}
where the overdot $\dot{\;}$ is shorthand for derivatives with respect to $t$, and $D_a$ is a spatial $SO(3)$ covariant derivative acting as $D_aB^i=\partial_a B^i+{\epsilon^i}_{jk}A^j_a B^k$. The final term in Eq.~(\ref{decomposition}) gives a boundary term after integration over $\mathcal{S}$, which we will assume vanishes. The action (\ref{Plebanskiaction}) then becomes the {\em extended canonical Pleba\'nski action}
\begin{align}
S_{{\rm xC}} &=\;\frac{1}{8\sqrt{\sigma}\,\pi G}\int {\rm d}^4 x\bigl[\dot{A}^i_a\tilde{E}_i^a + \alpha^i\,D_a\tilde{E}^a_i-\tilde{N}\left({\rm tr}\,M-\Lambda\right)\nonumber
\\ & \;\quad-V^a\Bigl(F^i_{ab}\tilde{E}^b_i-A^i_a D_b\tilde{E}^b_i\Bigr)\nonumber
\\ & \;\quad-\tilde\varphi_{ij}\Bigl(\frac{F^{(i}_{ab}\epsilon^{j)kl}\tilde{E}^a_k\tilde{E}^b_l}{2\det\tilde{E}}-M^{ij}\Bigr)\bigr]\,.
\label{plebanskicanonical}
\end{align}
The densitised triad now appears as the canonically conjugate variable to the $SO(3)$ connection $A^i_a$, as expected. The fields $\alpha^i$, $V^a$ and $\tilde{N}$ are most naturally interpreted as Lagrange multipliers since the action does not depend on their time derivatives. In addition, we have the fields $\tilde\varphi_{ij}$ and $M^{ij}$ which are likewise non-dynamical. Including these fields into the canonical formulation would lead to a large phase space with second class constraints, since we would need to give these fields conjugate momenta. It is more convenient to eliminate these fields directly at the level of the action, by substituting $\frac{F^{(i}_{ab}\epsilon^{j)kl}\tilde{E}^a_k\tilde{E}^b_l}{2\det\tilde{E}}=M^{ij}$ back into the action. We obtain
\begin{align}
S_{{\rm Can}} &=\;\frac{1}{8\sqrt{\sigma}\,\pi G}\int {\rm d}^4 x\bigl[\dot{A}^i_a\tilde{E}_i^a + \alpha^i\,D_a\tilde{E}^a_i \label{plebanskicanonical2}
\\ & \;-V^a\left(F^i_{ab}\tilde{E}^b_i-A^i_a D_b\tilde{E}^b_i\right)-\tilde{N}\left(\frac{{\epsilon^{ij}}_k F^k_{ab}\tilde{E}^a_i\tilde{E}^b_j}{2\det\tilde{E}}-\Lambda\right)\bigr]\,, \nonumber
\end{align}
which yields the same field equations as the extended canonical action (\ref{plebanskicanonical}) for the remaining variables. Notice that one of the field equations coming from Eq.~(\ref{plebanskicanonical}) would be $\tilde\varphi_{ij}=\tilde{N}\delta_{ij}$. We can view this equation as a restriction of our ansatz for $\Sigma^i_{0a}$, which now reads 
\ben
\Sigma^i_{0a} = -\utilde{\epsilon}_{abc} V^b \tilde{E}^{ic} - \tilde{N}\utilde{E}^i_a\,.
\label{newsigma0a}
\een
This restricted ansatz is a parametrisation of solutions to the Lagrangian field equation $\Sigma^i\wedge\Sigma^j = \delta^{ij}\omega$; see Ref.~\cite{selfdual2forms} for a similar treatment. We can use the {\em canonical Pleba\'nski action} (\ref{plebanskicanonical2}) as a starting point for  a canonical formulation of Pleba\'nski gravity.  It is now clear that this action encodes exactly the Hamiltonian formulation of general relativity in terms of Ashtekar variables: we have a fundamental Poisson bracket
\ben
\{ A_a^i(x), \tilde{E}^b_j(x') \} =  8\sqrt{\sigma}\,\pi G \, \delta^i_j \delta^b_a \delta^{(3)}(x-x')
\een
and local constraints
\begin{align}
\tilde{\mathcal{G}}_i & = -D_a\tilde{E}^a_i \approx 0\,,
\label{gaussc}
\\ \tilde{\mathcal{D}}_a & = F^i_{ab} \tilde{E}^b_i - A^i_a D_b \tilde{E}^b_i \approx 0 \,,
\label{diffeoc}
\\ \mathcal{H} & = \frac{{\epsilon^{ij}}_k F^k_{ab}\tilde{E}^a_i\tilde{E}^b_j}{2\det\tilde{E}}-\Lambda \approx 0\,.
\label{hamiltonc}
\end{align}
These are the usual Gauss, diffeomorphism and Hamiltonian constraints of the Ashtekar formulation. Their smeared versions
\begin{align}
\mathcal{G}(\alpha) & = \frac{1}{8\sqrt{\sigma}\,\pi G} \int {\rm d}^3 x\,\alpha^i\tilde{\mathcal{G}}_i\,,\nonumber
\\\mathcal{D}({\bf V}) & = \frac{1}{8\sqrt{\sigma}\,\pi G} \int {\rm d}^3 x\,V^a\tilde{\mathcal{D}}_a\,,\nonumber
\\\mathcal{H}(\tilde{N}) & = \frac{1}{8\sqrt{\sigma}\,\pi G} \int {\rm d}^3 x\,\tilde{N}\mathcal{H}
\end{align}
satisfy a closed (first class) Poisson algebra
\begin{align}
\{\mathcal{G}(\alpha),\mathcal{G}(\beta)\} & = - \mathcal{G}([\alpha,\beta])\,,\nonumber
\\ \{\mathcal{G}(\alpha),\mathcal{D}({\bf V})\}&=-\mathcal{G}(\mathcal{L}_{\bf V}\alpha)\,,\nonumber
\\ \{\mathcal{G}(\alpha),\mathcal{H}(\tilde{N})\} & = 0\,,\nonumber
\\ \{\mathcal{D}({\bf U}),\mathcal{D}({\bf V})\} &= \mathcal{D}([{\bf U},{\bf V}])\,,\nonumber
\\ \{\mathcal{D}({\bf V}),\mathcal{H}(\tilde{N})\} & = \mathcal{H}(\mathcal{L}_{\bf V}\tilde{N}) \,,\nonumber
\\ \{\mathcal{H}(\tilde{N}_1),\mathcal{H}(\tilde{N}_2)\} & = \mathcal{D}^0 ({\bf X}(\tilde{N}_1,\tilde{N}_2))
\label{poissonalgebra}
\end{align}
where $\mathcal{L}_{\bf V}$ denotes the Lie derivative along the vector field $V^a$ and for the last line we define
\begin{align}
\mathcal{D}^0({\bf V}) & = \frac{1}{8\sqrt{\sigma}\,\pi G} \int {\rm d}^3 x\,V^a F^i_{ab}\tilde{E}^b_i\,,\nonumber
\\ X^a(\tilde{N}_1,\tilde{N}_2) & = \frac{\tilde{E}^a_i \tilde{E}^{ib}(\tilde{N}_2\partial_b\tilde{N}_1-\tilde{N}_1\partial_b\tilde{N}_2)}{(\det\tilde{E})^2}\,.
\label{d0vdef}
\end{align}
Note that $\mathcal{D}^0({\bf V})$ is not a new constraint but a linear combination of the Gauss and diffeomorphism constraints. In many treatments in the literature $\mathcal{D}^0({\bf V})$ rather than $\tilde{\mathcal{D}}_a$ is defined as the diffeomorphism constraint but, as we can already see from the algebra, it is $\tilde{\mathcal{D}}_a$ which generates spatial diffeomorphisms via Lie derivative along a vector field (see, e.g., Ref.~\cite{Giulini} for a similar discussion). Concretely, we have $\delta_{\bf V}\mathcal{O}=\{\mathcal{O},\mathcal{D}({\bf V})\}$ with
\begin{align}
\delta_{\bf V}A^i_a & = V^b F^i_{ba}+D_a(V^bA_b^i)\,,\nonumber
\\ \delta_{\bf V}\tilde{E}^a_i&=\partial_b(V^b\tilde{E}^a_i)-\tilde{E}^b_i\partial_b V^a
\end{align}
which is exactly the action of the Lie derivative on these variables.  Likewise one can check that the Gauss constraint generates local $SO(3)$ gauge transformations via $\delta_\alpha\mathcal{O}=\{\mathcal{O},\mathcal{G}(\alpha)\}$. $\mathcal{H}$ generates transformations corresponding to reparametrisations of the embedding parameter $t$, so that together with $\tilde{\mathcal{D}}_a$ one obtains the full diffeomorphism group.

The total Hamiltonian is now given by
\ben
H_{{\rm Ple}} = \mathcal{G}(\alpha) + \mathcal{D}({\bf V}) + \mathcal{H}(\tilde{N})
\label{hple}
\een
which depends on the unconstrained fields $\alpha^i$, $V^a$ and $\tilde{N}$, whose choice determines a gauge used in the evolution. 

\subsection{Reality conditions and the metric}
\label{realitysec}

So far, this defines a canonical formulation for either Euclidean general relativity ($\sigma=+1$) or for complex general relativity ($\sigma=-1$). To identify Lorentzian solutions in the latter case, we need to impose reality conditions. In the Pleba\'nski formulation these are given by
\ben
\Sigma^i\wedge\overline{\Sigma^j} = 0\,,\quad{\rm Re}(\Sigma_i\wedge\Sigma^i)=0\,.
\een
Notice that so far, we could work in a holomorphic formalism: the Lagrangian was holomorphic in the fields, and the Hamiltonian theory can be defined in terms of a holomorphic Poisson bracket. These properties are now broken by the reality conditions, which involve $\Sigma^i$ and its complex conjugate rather than $\Sigma^i$ alone. In practice, we treat the reality conditions as additional, non-holomorphic constraints on initial data, whose consistency under time evolution must be checked; see Ref.~\cite{Sahlmann} for an analysis of this approach in Ashtekar variables. 

Using Eqs.~(\ref{sigmadecompose}), (\ref{denstriad}) and (\ref{newsigma0a}), we can write the 2-forms $\Sigma^i$ as
\ben
\Sigma^i = -\tilde{N}\,\utilde{E}^i_a \,{\rm d}t\wedge w^a + \half \tilde{E}^{ia}\utilde{\epsilon}_{abc}\,w^b\wedge w^c
\label{newsigmadec}
\een
with $w^a=V^a {\rm d}t+{\rm d}x^a$, and the reality conditions become
\begin{align}
\tilde{N}\left(\tilde{E}^{ia}\overline{\utilde{E}^j_a}-\overline{\tilde{E}^{ia}}\utilde{E}^j_a\right) & = \utilde{\epsilon}_{abc}\tilde{E}^{ib}\overline{\tilde{E}^{jc}}\left(V^a-\overline{V^a}\right)\,,\nonumber
\\{\rm Re}\,\tilde{N} & = 0\,.
\label{realitycond}
\end{align}
These conditions can be understood in terms of the tetrad derived from the $\Sigma^i$: the tetrad $e^I$ defined by
\ben
e^0 = \frac{\tilde{N}}{\sqrt{\det\tilde E}}\,{\rm d}t\,,\quad e^i = {\rm i}\sqrt{\det\tilde{E}} \, \utilde{E}^i_a\,w^a
\een
indeed satisfies Eq.~(\ref{tetracond}) with Eq.~(\ref{newsigmadec}). The associated (Urbantke) metric is then defined by
\begin{align}
g & = -\frac{\tilde{N}^2}{\det\tilde E}\,{\rm d}t\otimes{\rm d}t - (\det\tilde{E})\utilde{E}^i_a \utilde{E}_{ib}\,w^a\otimes w^b\,.
\label{metric}
\end{align}
We can see that when compared to the usual Arnowitt--Deser--Misner canonical decomposition of the metric, $\tilde{N}$ is a rescaled or densitised lapse function, and $V^a$ encodes the shift vector. The reality conditions tell us that $\tilde{N}$ should be purely imaginary; if we then choose $V^a$ to be real-valued, we find a second reality condition
\ben
{\rm Im} \left(\utilde{E}^i_a \utilde{E}_{ib}\right) = 0\,.
\een
This also implies that ${\rm Im}((\det\tilde{E})^2)=0$ and $\det\tilde{E}$ must either be real-valued or purely imaginary, and there are four separate solution sectors to the reality conditions: the metric (\ref{metric}) can be real Lorentzian with signature $(-+++)$ or $(+---)$ or it can be of the form ${\rm i}$ times a Lorentzian metric. This ambiguity in the solutions to the reality conditions in Pleba\'nski gravity is known \cite{Krasnovbook}, and it is discussed in the context of cosmological models in Ref.~\cite{ConnCosmo}; one could see it as a feature which allows for dynamical signature change in the metric (see also Ref.~\cite{signchange}). Once we require the fields to solve all field equations, the field equation ${\rm tr}\,M=\Lambda$, for real $\Lambda$,  would exclude the imaginary metric sectors. In the next section we will consider  unimodular extensions in which $\Lambda$ is an integration constant which is in general complex, and so these imaginary sectors are allowed by the theory \cite{UnimPleb}.

This way of imposing reality conditions is rather different from the usual treatment in the Ashtekar formalism, where $\tilde{E}^a_i$ itself is required to be real, perhaps up to local complex $SO(3)$ rotations \cite{Sahlmann}.

In the Euclidean formulation, reality conditions are not needed and the reconstructed metric is of the same form (\ref{metric}) and now always real Euclidean, but for $\det\tilde{E}>0$ it would be negative definite, so the signature ambiguity persists even in this case.

\subsection{First order and pure connection theories}

As we mentioned above, the Pleba\'nski action (\ref{Plebanskiaction}) can be used as a starting point for equivalent formulations in terms of fewer independent variables. In particular, substituting $\Sigma^i=(M^{-1})_{ij}F^j$ into Eq.~(\ref{Plebanskiaction}) yields a first-order action
\ben
S_{{\rm FO}}[A,M,\omega]=\frac{1}{16\sqrt{\sigma}\pi G}\int \bigl[M^{-1}_{ij}F^i\wedge F^j + \omega\left({\rm tr}\, M-\Lambda\right)\bigr]\,,
\een
which could also be used as a starting point for a canonical analysis. Decomposing the connection $A^i$ as in Eq.~(\ref{sigmadecompose}) one now obtains immediately
\begin{align}
S_{{\rm FO}} &=\;\frac{1}{8\sqrt{\sigma}\,\pi G}\int {\rm d}^4 x\bigl[\half\dot{A}^i_a M^{-1}_{ij}F^i_{bc}\tilde{\epsilon}^{abc} -\tilde{N}\left({\rm tr}\,M-\Lambda\right)\nonumber
\\ & \quad +\half A_0^i\,D_a\left(M^{-1}_{ij}F^i_{bc}\tilde\epsilon^{abc}\right)\bigr]\,.
\end{align}
This contains fewer independent variables than Eq.~(\ref{plebanskicanonical}) but the canonical phase space structure is not immediately apparent. From the first term in the action, if we introduce canonical momenta $\tilde{E}_i^a$ for the connection we obtain a primary constraint
\ben
\tilde{E}_i^a - \half M^{-1}_{ij}F^i_{bc}\tilde{\epsilon}^{abc} \approx 0
\een
which, using the inverse triad (\ref{invtriad}), can be written as
\ben
M^{ij} \approx \frac{F^i_{bc}\epsilon^{jkl}\tilde{E}^b_k\tilde{E}^c_l}{2\det\tilde E}\,.
\een
This constraint splits into a symmetric and an antisymmetric part, and given the symmetry of $M^{ij}$ we find $F^i_{ab}\tilde{E}^b_i\approx 0$. These constraints are precisely the additional constraints appearing in the extended canonical action (\ref{plebanskicanonical}), and the theories are seen to be equivalent.

In short, the transition to the canonical theory requires us to reintroduce the $\Sigma^i$ fields, which were integrated out to derive the first order theory, as canonical momenta to the connection. The same is true in a pure connection theory where one also eliminates the matrix $M^{ij}$ at the Lagrangian level. The canonical analysis of the pure connection formalism for standard (non-unimodular) general relativity can be found in Ref.~\cite{celda}. The intermediate first order theory was also used in the canonical analysis of Ref.~\cite{Smolin}. At the Hamiltonian level, using these actions seems to give no new insights compared to the original Pleba\'nski formulation, although one can obtain the same results more directly.

\section{Canonical unimodular Pleba\'nski gravity}
\label{canonsec2}

We can now extend the results of the canonical analysis of standard Pleba\'nski gravity to the unimodular versions proposed in Eqs.~(\ref{preferredvolume}) and (\ref{plebanskiht}).  The analysis of the preferred-volume theory is new, whereas the analysis of the parametrised Henneaux--Teitelboim version has been partially discussed in Ref.~\cite{Smolin}.

\subsection{Preferred-volume theory}

Starting with the preferred-volume theory defined by Eq.~(\ref{preferredvolume}), we note that it differs from the Pleba\'nski action (\ref{Plebanskiaction}) by the absence of a $\Lambda$ term and by the replacement of the dynamical field $\omega$ by a background field $\omega_0$, which can be written as $\omega_0=-2\tilde{N}_0\,{\rm d}^4 x$ for some background scalar density $\tilde{N}_0$. The other fields in the theory $(A^i,\Sigma^i,M^{ij})$ are the same  as in the conventional Pleba\'nski formalism, so most of the steps of the previous section go through with very minor modifications.

In analogy with Eq.~(\ref{plebanskicanonical}), we can define an {\em extended canonical preferred-volume action}
\begin{align}
S_{{\rm xCPV}} &=\;\frac{1}{8\sqrt{\sigma}\,\pi G}\int {\rm d}^4 x\bigl[\dot{A}^i_a\tilde{E}_i^a + \alpha^i\,D_a\tilde{E}^a_i-\tilde{N}_0\,{\rm tr}\,M\nonumber
\\ & \;\quad-V^a\Bigl(F^i_{ab}\tilde{E}^b_i-A^i_a D_b\tilde{E}^b_i\Bigr)\nonumber
\\ & \;\quad-\tilde\varphi_{ij}\Bigl(\frac{F^{(i}_{ab}\epsilon^{j)kl}\tilde{E}^a_k\tilde{E}^b_l}{2\det\tilde{E}}-M^{ij}\Bigr)\bigr]\,.
\end{align}
where we have only replaced $\tilde{N}\rightarrow\tilde{N}_0$ and $\Lambda\rightarrow 0$ compared to Eq.~(\ref{plebanskicanonical}). As with the non-unimodular version, we can fix $\tilde\varphi_{ij}=\tilde{N}_0 \delta_{ij}$ and replace $M^{ij}=\frac{F^{(i}_{ab}\epsilon^{j)kl}\tilde{E}^a_k\tilde{E}^b_l}{2\det\tilde{E}}$ to eliminate the redundant fields $\tilde\varphi_{ij}$ and $M^{ij}$, and find a {\em canonical preferred-volume action}
\begin{align}
S_{{\rm CanPV}} &=\;\frac{1}{8\sqrt{\sigma}\,\pi G}\int {\rm d}^4 x\bigl[\dot{A}^i_a\tilde{E}_i^a + \alpha^i\,D_a\tilde{E}^a_i
\\ & \;-V^a\left(F^i_{ab}\tilde{E}^b_i-A^i_a D_b\tilde{E}^b_i\right)-\tilde{N}_0\,\frac{{\epsilon^{ij}}_k F^k_{ab}\tilde{E}^a_i\tilde{E}^b_j}{2\det\tilde{E}}\bigr]\,. \nonumber
\end{align}
There is now a difference between the canonical formulation of this preferred-volume theory and the previously discussed canonical Pleba\'nski formulation: since $\tilde{N}_0$ is not a Lagrange multiplier but a fixed background field, we do not initially have a Hamiltonian constraint, but only the Gauss constraints $\tilde{\mathcal{G}}_i$ and diffeomorphism constraints $\tilde{\mathcal{D}}_a$ which take the same form as in Eqs.~(\ref{gaussc}) and (\ref{diffeoc}).

The total (naive) Hamiltonian is given by
\ben
H^{(0)} = \mathcal{G}(\alpha) + \mathcal{D}({\bf V}) + \frac{1}{8\sqrt{\sigma}\,\pi G}\int {\rm d}^3 x\,\tilde{N}_0\,\frac{{\epsilon^{ij}}_k F^k_{ab}\tilde{E}^a_i\tilde{E}^b_j}{2\det\tilde{E}}
\een
which includes a non-constraint part, hence the time evolution it generates is not pure gauge. The non-constraint part of this Hamiltonian coming from the integral term in the above expansion has a non-weakly vanishing Poisson bracket with the diffeomorphism constraint, and the consistency condition $\{ \mathcal{D}({\bf U}), H^{(0)} \} \approx 0$ yields a secondary constraint on the variables given by
\ben
\mathcal{K}_a = \partial_a \left( \frac{{\epsilon^{ij}}_k F^k_{ab}\tilde{E}^a_i\tilde{E}^b_j}{2\det\tilde{E}} \right)\,,
\een
which is the spatial gradient of the Hamiltonian constraint $\mathcal{H}$ encountered in the non-unimodular version in Eq.~(\ref{hamiltonc}). We can define its smeared form by
\ben
\mathcal{K}(\tilde{\bf T}) = \frac{1}{8\sqrt{\sigma}\,\pi G} \int {\rm d}^3 x\,\tilde{T}^a\mathcal{K}_a
\een
where $\tilde{T}^a$ is a weight $+1$ vector density. This new constraint is first class with the Gauss and diffeomorphism constraints as
\begin{align}
 \{\mathcal{G}(\alpha),\mathcal{K}(\tilde{{\bf T}})\} & = 0\,,\nonumber
\\ \{\mathcal{D}({\bf V}),\mathcal{K}(\tilde{{\bf T}})\} & = \mathcal{K}(\mathcal{L}_{\bf V}\tilde{{\bf T}}) \,,\nonumber
\\ \{\mathcal{K}(\tilde{{\bf T}}),\mathcal{K}(\tilde{{\bf L}})\} & = \mathcal{D}^0 ({\bf X}(\partial_a\tilde{T}^a,\partial_b\tilde{L}^b))
\end{align}
with $\mathcal{D}^0$ and $X^a$ defined as in Eq.~(\ref{d0vdef}). We must add a term $\mathcal{K}(\tilde{\bf T})$ to the naive Hamiltonian $H^{(0)}$ to obtain the full Hamiltonian $H_{{\rm PV}}$ of this theory, which defines the most general consistent time evolution:
\ben
H_{{\rm PV}} = H^{(0)} + \mathcal{K}(\tilde{\bf T})\,.
\een

We already understand the nature of the gauge transformations generated by the Gauss and diffeomorphism constraints. To understand the nature of the gauge transformations generated by the new constraint $\mathcal{K}$, notice that $\mathcal{K}(\tilde{{\bf T}})=\mathcal{H}(-\partial_a\tilde{T}^a)$ and hence for any $\mathcal{O}$
\ben
\{ \mathcal{O}, \mathcal{K}(\tilde{{\bf T}}) \} = -\{\mathcal{O}, \mathcal{H}(\partial_a\tilde{T}^a) \}\,.
\een
We see the gauge transformations generated from $\mathcal{K}$ are the same as the ones generated by $\mathcal{H}$ in the non-unimodular theory, except that the gauge parameter $\tilde{N}$ is restricted to be a total divergence $-\partial_a\tilde{T}^a$. These are a restricted set of time reparametrisation transformations, corresponding to the reduced symmetry from all diffeomorphisms to volume-preserving diffeomorphisms.

In the complex version of the theory, the tetrad and metric associated to the 2-forms $\Sigma^i$ are now obtained as
\begin{align}
e^0 & = \frac{\tilde{N}_0}{\sqrt{\det\tilde E}}\,{\rm d}t\,,\quad e^i = {\rm i}\sqrt{\det\tilde{E}} \, \utilde{E}^i_a\,w^a\,,
\\g & = -\frac{\tilde{N}_0^2}{\det\tilde E}\,{\rm d}t\otimes{\rm d}t - (\det\tilde{E})\utilde{E}^i_a \utilde{E}_{ib}\,w^a\otimes w^b\,.
\end{align}
In particular, one finds that $\sqrt{|g|}=|\tilde{N}_0|$, as expected. The reality conditions take the same form as in Eq.~(\ref{realitycond}), except that ${\rm Re}\,\tilde{N}_0 = 0$ now refers to the background field $\tilde{N}_0$. Hence Lorentzian solutions only exist when the background scalar density $\tilde{N}_0$ is purely imaginary. In contrast, Euclidean signature solutions only exist when $\tilde{N}_0$ is real valued. In this case, one takes all of the dynamical fields and Lagrange multipliers to be real also.

While this theory does not have the same Hamiltonian constraint as standard Pleba\'nski gravity, the constraint $\mathcal{K}_a$ implies that $\mathcal{H}$ is constant on each constant-time hypersurface. A quick calculation shows that we also have
\ben
\dot{\mathcal{H}} = \{\mathcal{H},H_{{\rm PV}}\} \approx 0\,.
\een
Hence, we see that the constraint (\ref{hamiltonc}) is replaced by a version in which $\Lambda$ is a free integration constant of the theory, as expected in unimodular formulations of gravity. Notice that this integration constant can in general be complex, and even if reality conditions are imposed can be purely imaginary. One could impose an additional reality condition to exclude these solutions \cite{UnimPleb}.

\subsection{Parametrised theory}

Here the starting point is the parametrised action defined {\em \`a la} Henneaux--Teitelboim, Eq.~(\ref{plebanskiht}). It contains a new dynamical variable $T$, which can be expanded as
\begin{align}
T & = \,\tilde\tau\,w^1\wedge w^2\wedge w^3 -\half \tilde{T}^a \utilde{\epsilon}_{abc}\, {\rm d}t \wedge w^b \wedge w^c\nonumber
\\ & = \,\tilde\tau\,{\rm d}^3 x -\half\left(\tilde{T}^a-\tilde\tau\,V^a\right)\utilde{\epsilon}_{abc}\, {\rm d}t \wedge {\rm d}x^b \wedge {\rm d}x^c
\label{Tparam}
\end{align}
where the 1-forms $w^a=V^a {\rm d}t+{\rm d}x^a$ were introduced below Eq.~(\ref{newsigmadec}), $\tilde\tau$ is a weight $+1$ scalar density and $\tilde{T}^a$ is a weight $+1$ vector density. The exterior derivative of the 3-form $T$ is computed to be
\ben
{\rm d}T = {\rm d}^4 x\,\left[\dot{\tilde\tau}+\partial_a\left(\tilde{T}^a-\tilde\tau V^a\right)\right]\,.
\label{dTexp}
\een
With the other variables defined as in Section~\ref{canonsec1}, we see that the action (\ref{plebanskiht}) becomes
\begin{align}
S_{{\rm xCHT}} &=\;\frac{1}{8\sqrt{\sigma}\,\pi G}\int {\rm d}^4 x\bigl[\dot{A}^i_a\tilde{E}_i^a + \dot{\tilde\tau}\,{\rm tr}\,M + \alpha^i\,D_a\tilde{E}^a_i\nonumber
\\ & \;\;-V^a\Bigl(F^i_{ab}\tilde{E}^b_i-A^i_a D_b\tilde{E}^b_i-\tilde\tau\,\partial_a{\rm tr}\,M\Bigr)
\\ & \;\;-\tilde\varphi_{ij}\Bigl(\frac{F^{(i}_{ab}\epsilon^{j)kl}\tilde{E}^a_k\tilde{E}^b_l}{2\det\tilde{E}}-M^{ij}\Bigr)-\tilde{T}^a\partial_a{\rm tr}\,M\bigr]\,.\nonumber
\end{align}
On inspection, we see an extra term $\dot{\tilde\tau}\,{\rm tr}\,M$ which contributes to the symplectic part of the action; $\tilde\tau$ and ${\rm tr}\, M$ are now canonically conjugate. To make this explicit we can decompose $M^{ij}=\Psi^{ij}+\frac{1}{3}\lambda\delta^{ij}$ where $\Psi^{ij}$ is the trace-free part and $\lambda={\rm tr}\,M$. We can also decompose $\tilde\varphi_{ij}=\tilde\chi_{ij}+\tilde{N}\delta_{ij}$ in a similar fashion. We then have
\begin{align}
& \tilde\varphi_{ij}\Bigl(\frac{F^{(i}_{ab}\epsilon^{j)kl}\tilde{E}^a_k\tilde{E}^b_l}{2\det\tilde{E}}-M^{ij}\Bigr) 
\\ = & \,\tilde\chi_{ij}\Bigl(\frac{F^{(i}_{ab}\epsilon^{j)kl}\tilde{E}^a_k\tilde{E}^b_l}{2\det\tilde{E}}\bigr|_{{\rm tf}}-\Psi^{ij}\Bigr) + \tilde{N}\left(\frac{F^{i}_{ab}{\epsilon_i}^{kl}\tilde{E}^a_k\tilde{E}^b_l}{2\det\tilde{E}}-\lambda\right) \nonumber
\end{align}
where $\bigr|_{{\rm tf}}$ denotes the trace-free part, explicitly
\ben
\frac{F^{(i}_{ab}\epsilon^{j)kl}\tilde{E}^a_k\tilde{E}^b_l}{2\det\tilde{E}}\bigr|_{{\rm tf}}=\frac{F^{(i}_{ab}\epsilon^{j)kl}\tilde{E}^a_k\tilde{E}^b_l}{2\det\tilde{E}}-\frac{1}{3}\delta^{ij}\frac{F^{m}_{ab}{\epsilon_m}^{kl}\tilde{E}^a_k\tilde{E}^b_l}{2\det\tilde{E}}\,.
\een
In the non-unimodular Pleba\'nski formalism and in the preferred-volume unimodular theory, we were able to eliminate the fields $M^{ij}$ and $\tilde\varphi_{ij}$ in their entirety. In this case, we may only eliminate the trace-free parts $\Psi^{ij}$ and $\tilde\chi_{ij}$ since the trace parts are now dynamical. Removing the redundant trace-free variables, we obtain the {\em canonical parametrised unimodular Pleba\'nski action}
\begin{align}
S_{{\rm CanHT}} &=\;\frac{1}{8\sqrt{\sigma}\,\pi G}\int {\rm d}^4 x\bigl[\dot{A}^i_a\tilde{E}_i^a + \dot{\tilde\tau}\lambda + \alpha^i\,D_a\tilde{E}^a_i\nonumber
\\ & \;\;-V^a\Bigl(F^i_{ab}\tilde{E}^b_i-A^i_a D_b\tilde{E}^b_i-\tilde\tau\,\partial_a\lambda\Bigr)\nonumber
\\ & \;\;-\tilde{N}\Bigl(\frac{{\epsilon^{ij}}_kF^{k}_{ab}\tilde{E}^a_i\tilde{E}^b_j}{2\det\tilde{E}}-\lambda\Bigr)-\tilde{T}^a\partial_a\lambda\bigr]\,.
\end{align}
The transformation into the Hamiltonian setting is then clear. We have dynamical fields $A_a^i$, $\tilde{E}_i^a$, $\tilde\tau$, $\lambda$ with 
\begin{align}
\{ A_a^i(x), \tilde{E}^b_j(x') \} & =  8\sqrt{\sigma}\,\pi G \, \delta^i_j \delta^b_a \delta^{(3)}(x-x')\,,\nonumber
\\ \{ \tilde\tau(x), \lambda(x') \} & =  8\sqrt{\sigma}\,\pi G \, \delta^{(3)}(x-x')\,;\nonumber
\end{align}
the fields $\alpha^i$, $V^a$, $\tilde{N}$ and $\tilde{T}^a$ are Lagrange multipliers enforcing constraints
\begin{align}
\tilde{\mathcal{G}}_i & = -D_a\tilde{E}^a_i \approx 0\,,
\\ \tilde{\mathcal{D}}'_a & = F^i_{ab} \tilde{E}^b_i - A^i_a D_b \tilde{E}^b_i -\tilde\tau\,\partial_a\lambda\approx 0 \,,
\\ \mathcal{H}' & = \frac{{\epsilon^{ij}}_k F^k_{ab}\tilde{E}^a_i\tilde{E}^b_j}{2\det\tilde{E}}-\lambda \approx 0\,,
\\ \mathcal{J}_a & = \partial_a\lambda\approx 0\,.
\end{align}
The Gauss constraint takes the same form (\ref{gaussc}) as in the usual Pleba\'nski theory, whereas the diffeomorphism constraint $\tilde{\mathcal{D}}'_a$ picks up an additional term compared to Eq.~(\ref{diffeoc}). This corresponds to the fact that the new fields $\lambda$ and $\tilde\tau$ transform nontrivially (as a scalar and as a scalar density) under spatial diffeomorphisms. The new Hamiltonian constraint $\mathcal{H}'$ shows that the dynamical variable $\lambda$ replaces the cosmological constant in usual Pleba\'nski gravity. Finally, the new constraint $\mathcal{J}_a$ forces $\lambda$ to be constant on each spatial hypersurface. 

The constraints $\tilde{\mathcal{G}}_i, \tilde{\mathcal{D}}'_a$ and $\mathcal{H}'$ satisfy the same Poisson algebra as their non-unimodular counterparts, given in Eq.~(\ref{poissonalgebra}). We may define a smeared version of $\mathcal{J}_a$ by
\ben
\mathcal{J}(\tilde{\bf T}) = \frac{1}{8\sqrt{\sigma}\,\pi G} \int {\rm d}^3 x\,\tilde{T}^a\mathcal{J}_a\,;
\een
then the only nonvanishing Poisson bracket of $\mathcal{J}$ is with the diffeomorphism constraint,
\ben
\\ \{\mathcal{D}'({\bf V}),\mathcal{J}(\tilde{{\bf T}})\} = \mathcal{J}(\mathcal{L}_{\bf V}\tilde{{\bf T}}) \,.\nonumber
\een
Hence we have a first-class constraint algebra and no further constraints need to be added. The total Hamiltonian of the theory can be defined as
\ben
H_{{\rm HT}} = \mathcal{G}(\alpha) + \mathcal{D}'({\bf V}) + \mathcal{H}'(\tilde{N}) + \mathcal{J}(\tilde{\bf T})\,.
\een
The gauge transformations generated by these constraints correspond to local $SO(3)$ transformations, spatial diffeomorphisms and time reparametrisations as in the standard non-unimodular Pleba\'nski theory; the constraint $\mathcal{J}_a$ only generates transformations on $\tilde\tau$, namely
\ben
\delta_{\tilde{\bf L}}\tilde\tau=\{\tilde\tau,\mathcal{J}(\tilde{{\bf L}})\}=-\partial_a\tilde{L}^a\,.
\een
These transformations correspond to symmetries of the action (\ref{plebanskiht}) under $T\rightarrow T+\theta$ where $\theta$ is a closed 3-form satisfying ${\rm d}\theta = 0$, which may be seen as an additional gauge symmetry. To recover this symmetry in the Hamiltonian picture, we need to define the behaviour of the Lagrange multiplier field $\tilde{T}^a$ under these gauge transformations using the invariance of the action (see Appendix \ref{appendix}). For the transformations generated by $\mathcal{J}$, we find
\ben
\delta_{\tilde{\bf L}} \tilde{T}^a = \dot{\tilde{L}}^a - \mathcal{L}_{{\bf V}}\tilde{L}^a
\een
and hence
\begin{align}
\delta_{\tilde{\bf L}} T & = -\partial_a\tilde{L}^a\,{\rm d}^3 x \nonumber
\\ & \;\;-\half\left(\dot{\tilde{L}}^a - \partial_d(V^d \tilde{L}^a -\tilde{L}^d V^a)\right)\utilde{\epsilon}_{abc} {\rm d}t \wedge {\rm d}x^b \wedge {\rm d}x^c
\end{align}
using the parametrisation of $T$ defined in Eq.~(\ref{Tparam}). Evaluating the exterior derivative ${\rm d}\delta_{\tilde{\bf L}} T$ reveals that $\delta_{\tilde{\bf L}} T$ is closed, ${\rm d}\delta_{\tilde{\bf L}} T=0$. The same symmetry is discussed in the usual Henneaux--Teitelboim formulation of unimodular gravity in Ref.~\cite{HenneauxTeitelboim}.

The Hamiltonian equations of motion for the field $\lambda$ imply that $\dot\lambda=\{\lambda,H_{{\rm HT}}\}=0$ so that $\lambda$ is again constant in spacetime, not just on each spatial hypersurface.  

The discussion of reality conditions in this theory largely mirrors the one of Sec.~\ref{realitysec}, except that there is now also a dynamical field $\lambda$ which represents the cosmological constant in Einstein's equations. By imposing reality conditions on $\lambda$ we can exclude unwanted sectors of theory. In particular, demanding $\lambda$ to be real excludes the solutions with imaginary metric.

A significant property of the parametrised approach, whether in metric \cite{HenneauxTeitelboim} or connection variables \cite{Smolin}, is that the volume form is exact; the total volume of a portion of spacetime of the form $[t_0,t_1]\times\mathcal{S}$ can be written as a difference between two integrals over the boundary hypersurfaces. Hence these boundary integrals can be used to define a preferred time variable, the volume time, in this formulation of unimodular gravity.  

In the Lagrangian setting, this property can be seen in one of the field equations arising from the action (\ref{plebanskiht}),
\ben
\Sigma^i \wedge \Sigma^j = 2\delta^{ij}\,{\rm d}T\,,
\een
see also Ref.~\cite{UnimPleb} for more discussion on this. One can recover this result from the
canonical theory also. To see this, consider the equation of motion for $\tilde\tau$ computed via 
\ben
\dot{\tilde\tau} = \{ \tilde\tau, H_{{\rm HT}} \} = -\tilde{N} - \partial_a\left(\tilde{T}^a-\tilde\tau\,V^a\right)
\label{taueom}
\een
Hence, from Eq.~(\ref{dTexp}) we have ${\rm d}T = -\tilde{N}\,{\rm d}^4 x$. 

Now recall that the metric (\ref{metric}) associated to the variables of the Pleba\'nski theory has determinant $\tilde{N}^2$. The reality conditions in the Lorentzian theory imply that $\tilde{N}$ is imaginary, so that ${\rm d}T$ is also imaginary. In the Euclidean theory all fields are real. Hence, an exact volume form compatible with the metric is given by $\omega_{{\rm HT}}=\sqrt{\sigma}\,{\rm d}T$ for both Lorentzian and Euclidean solutions (the overall sign represents a choice of orientation, which cannot be determined from the metric alone). 

If we now define the volume time $t_{\rm Vol}$ between hypersurfaces $t=t_0$ and $t=t_1$ to be the spacetime volume of the region bounded by them, we find
\begin{align}
t_{\rm Vol} (t_0, t_1) & = \int_{ [t_0,t_1]\times\mathcal{S}} \omega_{{\rm HT}} \nonumber
\\ & = \sqrt{\sigma} \left(\int_{t=t_1} {\rm d}^3 x \, \tilde\tau -\int_{t=t_0} {\rm d}^3 x \, \tilde \tau \right)\,.
\end{align}

This provides a geometric interpretation for the field $\tilde\tau$ as encoding the volume time between constant $t$ hypersurfaces.
This volume time is not fully gauge invariant; the transformations on the variable $\tilde\tau$ generated from the constraint $\mathcal{H}'(\tilde{N})$ produce the time reparametrisation transformations on $t_{\rm Vol}$ that we might expect, namely
\begin{align}
\delta_{\tilde{N}} t_{\rm Vol}(t_0, t_1) & = - \sqrt{\sigma} \left(\int_{t=t_1} {\rm d}^3 x \, \tilde{N} -\int_{t=t_0} {\rm d}^3 x \, \tilde{N} \right)\nonumber
\\ & = \sqrt{\sigma} \left(\int_{t=t_1} {\rm d}^3 x \, \dot{\tilde\tau} -\int_{t=t_0} {\rm d}^3 x \, \dot{\tilde\tau} \right) 
\\ & = \frac{\partial t_{\rm Vol}}{\partial t_0} + \frac{\partial t_{\rm Vol}}{\partial t_1}\,.
\end{align}
In the first equality we use the shift $\delta_{\tilde{N}}\tilde\tau = \{\tilde \tau,\mathcal{H}'(N)\} = - \tilde{N}$, and in the second equality we use the equation of motion (\ref{taueom}). Gauge transformations generated from the other constraints have no effect on $t_{\rm Vol}$.
The clock function $t_{\rm Vol}$ satisfies
\ben
    t_{\rm Vol}(t_0, t_1) = t_{\rm Vol} (t_0, t_2) + t_{\rm Vol} (t_2, t_1)
\een
and it is monotonic with respect to the coordinate distance $|t_1-t_0|$ when $\int {\rm d}^3 x \,\tilde\tau$ is a monotonic function of coordinate time $t$, which corresponds to choosing $\tilde{N}$ such that $\sqrt{\sigma}\int {\rm d}^3 x \,\tilde N$ is either non-negative or non-positive. In standard general relativity we would usually assume that the lapse function is positive definite, which would be a similar restriction. The volume time is sensitive to the ordering of events, since $t_{\rm Vol} (t_0,t_1) = -t_{\rm Vol} (t_1,t_0)$.

\section{Conclusions}
\label{conclus}

The canonical structure of unimodular formulations of Pleba\'nski gravity largely mirrors what is seen in the conventional metric approach \cite{Moreorless,Unruh,kluson} or in connection variables \cite{connection}. In the preferred-volume version, we do not initially have a Hamiltonian constraint since the volume form is a fixed background field. Instead, a non-constraint part of the Hamiltonian generates ``true'' time evolution in the preferred time. Demanding that the constraints are preserved under time evolution then implies that the Hamiltonian density is equal to some undetermined integration constant, which replaces the cosmological constant of  general relativity. In a parametrised (Henneaux--Teitelboim) version of the theory, we have a new pair of canonically conjugate fields $\lambda$ and $\tilde\tau$. $\lambda$ is constrained by the equations of motion, and a new constraint, to be a spacetime constant, while $\tilde\tau$ can be used to construct a preferred volume time in the theory. In either approach, one adds only a single global degree of freedom compared to the degrees of freedom in the usual Pleba\'nski formalism for general relativity. Since we are working in a connection-based formulation closely related to Ashtekar variables, there is an additional Gauss constraint representing local $SO(3)$ transformations, but the interaction of this constraint with the others is straightforward to understand and as expected.

As we discussed at numerous points, additional complications arise from the need for reality conditions when we aim to find Lorentzian solutions in the complex theory (these are not discussed in Ref.~\cite{Smolin} which presents a shorter version of the canonical analysis for the parametrised theory). These subtleties are mostly known from the usual Pleba\'nski formalism, where they mean that the signature of the ``effective'' spacetime metric arising from the Urbantke construction cannot be fixed {\em a priori}, and the reality conditions even admit solutions for which the metric would be purely imaginary. In the standard Pleba\'nski approach, the latter would be excluded by the Einstein equations for real $\Lambda$, but here $\Lambda$ could be an arbitrary complex integration constant. Hence, to exclude imaginary metric solutions an additional reality condition would need to be imposed \cite{UnimPleb}. When attempting a canonical quantisation for instance with the methods of loop quantum gravity, as advocated in Ref.~\cite{Smolin}, this need for an additional reality condition would be an additional complication. One could focus on the Euclidean theory instead, where all fields are real and there is no need for reality conditions. Even in this case, one would have to accept solutions with negative definite metric. Hence, while there is a close connection between formulations of gravity based on self-dual 2-forms and the Ashtekar formalism, the fact that the densitised triad is not a fundamental variable in the Pleba\'nski theory does lead to differences between these approaches.

\acknowledgments
The work of SG is funded by the Royal Society through the University Research Fellowship Renewal URF$\backslash$R$\backslash$221005. 

\appendix
\section{Gauge transformations and Lagrange multipliers}
\label{appendix}
In the canonical formulation, local gauge transformations on the dynamical fields are generated symplectically from the first class constraints. However, the Poisson structure
does not prescribe gauge transformations for Lagrange multiplier fields as these are not phase space functions. It is possible to derive the gauge transformations of these Lagrange multiplier fields by requiring that the action should be invariant under arbitrary gauge transformations. This then connects with the transformation behaviour of these fields in the Lagrangian theory. Our discussion here follows Refs.~\cite{halliwell,gaugebook}.

Consider a general theory with first-class constraints generating gauge transformations, with action 
\ben
S[q,p,\lambda]=\int_{t_0}^{t_1} {\rm d}t \bigl[\dot{q}^i p_i - H(q,p) - \lambda^\alpha G_\alpha(q,p)\bigr]\,.
\een
Here $(q^i,p_i)$ are canonical coordinates and momenta, $H(q,p)$ is the non-constraint part of the Hamiltonian, $\lambda^\alpha$ are Lagrange multipliers and $G_\alpha$ are the constraints. We have written this action for a finite-dimensional mechanical system but the generalisation to field theory is straightforward.

Under a gauge transformation generated by some linear combination $\varepsilon^\beta G_\beta$ of the constraints, we have $\delta_\varepsilon f =\varepsilon^\beta \{f,G_\beta\}$ for any phase-space function $f$. Then the action $S$ changes as
\begin{align}
\delta_\varepsilon S & = \int_{t_0}^{t_1} {\rm d}t \bigl[p_i \delta_\varepsilon\dot{q}^i + \dot{q}^i \delta_\varepsilon p_i - \delta_\varepsilon H - G_\alpha\delta_\varepsilon\lambda^\alpha - \lambda^\alpha\delta_\varepsilon G_\alpha\bigr]\nonumber
\\& = \left[\varepsilon^\beta \frac{\partial G_\beta}{\partial p_i}p_i\right]_{t_0}^{t_1}\nonumber
\\& \;\; - \int_{t_0}^{t_1} {\rm d}t \bigl[ \varepsilon^\beta\dot{G}_\beta - \varepsilon^\beta\{ G_\beta, H+\lambda^\alpha G_\alpha\} + G_\alpha\delta_\varepsilon\lambda^a \bigr]
\nonumber
\\& = \left[\varepsilon^\beta \frac{\partial G_\beta}{\partial p_i}p_i-\varepsilon^\beta G_\beta\right]_{t_0}^{t_1}
\label{gaugetransf}
\\& \;\; + \int_{t_0}^{t_1} {\rm d}t \bigl[ \dot\varepsilon^\beta G_\beta + \varepsilon^\beta\{ G_\beta, H+\lambda^\alpha G_\alpha\} - G_\alpha\delta_\varepsilon\lambda^a \bigr]
\nonumber
\end{align}
where we use $\dot G_\beta=\dot{p}_i\{q^i,G_\beta\}-\dot{q}^i\{p_i,G_\beta\}$, $\delta_\varepsilon H =\varepsilon^\beta\{H,G_\beta\}$ and $\delta_\varepsilon G_\alpha =\varepsilon^\beta\{G_\alpha,G_\beta\}$ and we used two integrations by parts. If the gauge transformation parameters $\varepsilon^\beta$ are assumed to vanish at the initial and final times, the first line in Eq.~(\ref{gaugetransf}) vanishes. The change $\delta_\varepsilon\lambda^a$ in the Lagrange multiplier fields can then be defined by demanding that also the second line vanishes. Assuming the constraints $G_\alpha$ are independent, this gives as many independent equations as there are constraints.

We can now demonstrate this formalism for the canonical Pleba\'nski action (\ref{plebanskicanonical2}). Consider a smeared constraint $\Phi(\varepsilon)$ which can be a linear combination of the constraints (\ref{gaussc})-(\ref{hamiltonc}) with some Lagrange multiplier field $\varepsilon(t,x)$. We assume $\varepsilon(t_0,x)=\varepsilon(t_1,x)=0$. Then under a gauge transformation generated by $\Phi(\varepsilon)$, the action changes as
\begin{align}
\delta_\varepsilon S_{{\rm Can}} & = \int_{t_0}^{t_1} {\rm d}t \bigl[ \Phi(\dot\varepsilon) + \{ \Phi(\varepsilon), H_{{\rm Ple}}\} \nonumber
\\ & \qquad - \mathcal{G}(\delta_\varepsilon\alpha) - \mathcal{D}(\delta_\varepsilon{\bf V})-\mathcal{H}(\delta_\varepsilon\tilde{N}) \bigr]
\end{align}
where the total Pleba\'nski Hamiltonian $H_{{\rm Ple}}$ was defined in Eq.~(\ref{hple}). As an illustrative example, consider the transformation of the canonical action generated from the constraint $\tilde{\mathcal{D}}_a$ with gauge parameter $U^a$, which reads
\begin{align}
\delta_{\bf U} S_{{\rm Can}} & = \int_{t_0}^{t_1} {\rm d}t \bigl[ \mathcal{G}(\mathcal{L}_{{\bf U}}\alpha-\delta_{\bf U}\alpha) + \mathcal{H}(\mathcal{L}_{{\bf U}}\tilde{N}-\delta_{\bf U}\tilde{N}) \nonumber
\\&\qquad +\mathcal{D}(\dot{{\bf U}}+[{\bf U},{\bf V}]-\delta_{\bf U}{\bf V})  \bigr]\,.
\end{align}
By demanding that the action is invariant, we can read off the required transformation behaviour of the Lagrange multipliers under diffeomorphisms:
\ben
\delta_{\bf U}\alpha^i=\mathcal{L}_{{\bf U}}\alpha^i\,,\;\; \delta_{\bf U}\tilde{N}=\mathcal{L}_{{\bf U}}\tilde{N}\,,\;\; \delta_{\bf U}V^a=\dot{U}^a+[{\bf U},{\bf V}]^a\,.\nonumber
\een
These include the usual action via Lie derivative, but there is now a time-dependent piece $\dot{U}^a$ which takes care of the time-dependent gauge transformation. One could see $\delta_{\bf U}V^a$ as the spatial part of a commutator of spacetime vector fields $U^\mu$ and $V^\mu$ with $U^0 = 0$ and $V^0=-1$. 

The same formalism can be used to compute the transformation behaviour of Lagrange multiplier fields in all theories considered in this article \cite{Thesis}, and to identify Lagrangian symmetries at the Hamiltonian level.

\end{document}